\documentclass{emulateapj}
\usepackage{epsf}
\usepackage{epstopdf}
\usepackage{apjfonts}
\usepackage{xspace}
\usepackage{amsmath}
\usepackage{hyperref} 
\begin{document}
\shorttitle{Ellipticity Profiles of X-ray Clusters}
\submitted{The Astrophysical Journal, submitted}
\slugcomment{{\em The Astrophysical Journal, submitted}} 
\shortauthors{Lau et al.}

\title{Constraining Cluster Physics with the Shape of X-ray Clusters: \\ Comparison of Local X-ray Clusters vs. $\Lambda$CDM Clusters}

\author{Erwin T. Lau\altaffilmark{1,2},
Daisuke Nagai\altaffilmark{1,2},
Andrey V. Kravtsov\altaffilmark{3},  
Alexey Vikhlinin\altaffilmark{4,5},
Andrew R. Zentner\altaffilmark{6}
}



\altaffiltext{1}{Department of Physics, Yale University, New Haven,
 CT 06520, USA.}
\altaffiltext{2}{Yale Center for Astronomy and Astrophysics, Yale University, New Haven, CT 06520, USA; erwin.lau@yale.edu}
\altaffiltext{3}{Department of Astronomy \& Astrophysics, Kavli Institute for Cosmological Physics and Enrico Fermi Institute, 5640 South Ellis Ave., The University of Chicago, Chicago, IL 60637,USA.}
\altaffiltext{4}{Harvard-Smithsonian Center for Astrophysics, 60 Garden Street, Cambridge, MA 02138, USA.}
\altaffiltext{5}{Space Research Institute, 84/32 Profsojuznaya St., GSP-7, Moscow 117997, Russia}
\altaffiltext{6}{Department of Physics \& Astronomy and Pittsburgh Particle physics, Astrophysics, and Cosmology Center (PITT PACC), 
University of Pittsburgh, Pittsburgh, PA 15260, USA.}

\begin{abstract} 
Recent simulations of cluster formation have demonstrated that condensation of baryons into central galaxies during cluster formation can drive the shape of the gas distribution in galaxy clusters significantly rounder out to their virial radius.  These simulations generally predict stellar fractions within cluster virial radii that are $\sim 2-3$ times larger than the stellar masses deduced from observations.  In this paper we compare ellipticity profiles of simulated clusters performed with varying input  physics (radiative cooling, star formation, and supernova feedback), to the cluster ellipticity profiles derived from {\it Chandra} and {\it ROSAT} observations, in an effort to constrain the fraction of gas that cools and condenses into the central galaxies within clusters.  We find that local relaxed clusters have an average ellipticity of $\epsilon = 0.18\pm 0.05$ in the radial range of $0.04\leq r/r_{500}\leq 1$. At larger radii $r > 0.1r_{500}$, the observed ellipticity profiles agree well with the predictions of non-radiative simulations.  In contrast, the ellipticity profiles of simulated clusters that include dissipative gas physics deviate significantly from the observed ellipticity profiles at {\em all} radii.  The dissipative simulations overpredict (underpredict) ellipticity in the inner (outer) regions of galaxy clusters.  By comparing simulations with and without dissipative gas physics, we show that gas cooling causes the gas distribution to be more oblate in the central regions, but makes the outer gas distribution more spherical.  We find that late-time gas cooling and star formation are responsible for the significantly oblate gas distributions in cluster cores, but the gas shapes outside of cluster cores are set primarily by baryon dissipation at high redshift ($z\geq 2$). Our results indicate that the shapes of X-ray emitting gas in galaxy clusters, especially at large radii, can be used to place constraints on cluster gas physics, making it potential probes of the history of baryonic cooling in galaxy clusters. 

\end{abstract}


\section{Introduction}


Dark matter halos formed within the prevailing cold dark matter (CDM) paradigm  are expected to collapse from triaxial density peaks \citep{doroshkevich70,bardeen_etal86} and thus are generally expected to be triaxial \citep[e.g.,][]{frenk_etal88,dubinski_carlberg91,warren_etal92}.  Halo shapes can be approximated by an ellipsoid uniquely determined by three principal axes, with lengths and orientations that may vary as a function of distance to the halo center, as was demonstrated in detail by a number of numerical simulations \citep{cole_lacey96,thomas_etal98,jing_suto02,suwa_etal03,hopkins_etal05,kasun_evrard05,allgood_etal06, bett_etal07, gottloeber_yepes07,paz_etal08,lau_etal11,schneider_etal11}. The mean axis ratios for halos of a given mass are sensitive to cosmological parameters, particularly the normalization of the power spectrum \citep{ho_etal06,allgood_etal06}. Observational confirmation of the halo triaxiality predicted by simulations therefore serves as an important test of the CDM paradigm.  Furthermore, the shape of the intracluster medium (ICM) can also affect mass-observable scaling relations on the order of \ $\sim 10\%$ \citep[e.g.,][]{buote_humphrey11}. Hence understanding and constraining the triaxiality of the gas distribution is important for understanding sources of scatter in cluster mass-observable relations.
  
Gravitational lensing is a potentially powerful method to directly probe the shapes of halos. Recent analyses based on weak lensing \citep{evans_bridle09, oguri_etal10,oguri_etal11} and strong lensing \citep{richard_etal10,oguri_etal11} suggest that the shape of the total mass distribution is generally consistent with the predictions of CDM simulations. However, the lensing constraints on cluster shapes are often confined to the central regions of clusters (in the case of strong lensing), or are based on the assumption that the ellipticity is constant with radius inside the halo (in the case of weak lensing).  Stacking procedures to estimate halo shapes from shear signals require prior knowledge of the orientation of the axes of the halos, which introduces additional uncertainties in the estimated halo shapes when one assumes alignment between the observed galaxy distribution and the underlying dark matter \citep{bett11}. 

An alternative way to study halo shape is to measure the shape of the hot X-ray emitting gas in clusters under the assumption that gas is in hydrostatic equilibrium and thus follows the equipotential surfaces of the cluster \citep{binney_strimpel78,fabricant_etal84,buote_tsai95}. Given the relatively higher spatial resolution of X-ray maps compared to the lensing shear maps, and the fact that the measured ellipticity is independent of measured cluster mass and concentration, we can use gas shape to constrain the cluster ellipticity with higher accuracy. This can allow us to constrain the radial dependence of ellipticity predicted by numerical simulations \citep[e.g.][]{jing_suto02,allgood_etal06}.   Recent work suggests that the observed X-ray shape is consistent with the prediction of the CDM model \citep[e.g.,][]{flores_etal07,kawahara10}. 

Furthermore, the shape of dark matter distribution is quite sensitive to the degree of central concentration of mass. As baryons condense toward the center to form a central galaxy within a halo, the dark matter distribution becomes more spherical\citep{katz_gunn91,dubinski94,evrard_etal94,tissera_etal98,kazantzidis_etal04,hayashi_etal07,debattista_etal08,Tissera_etal09}. This effect increases with decreasing radius, but is substantial even at half of the virial radius \citep{kazantzidis_etal04}. The main mechanism behind this effect lies in adiabatic changes of the shapes of particle orbits in response to the more centrally-concentrated mass distribution after baryon dissipation \citep{dubinski94,debattista_etal08}. The change in the overall shape of the dark matter distribution makes the shape of the ICM rounder as well  \citep{lau_etal11}. Therefore, studies of the shape profile can  constrain the amount of cooling that has occurred during cluster formation. 

In \citet[hereafter Paper I]{lau_etal11}, we used a set of 16 high-resolution hydrodynamical cosmological cluster simulations with and without dissipative gas physics and showed that the shape and normalization of the X-ray ellipticity profile are sufficiently sensitive to distinguish between the shape of ICM predicted in simulations with and without cooling with high significance.  In particular, we showed that the baryon condensation  due to gas cooling  makes gas outside of cluster cores ($0.1\leq r/r_{500}\leq 1$), 
\footnote{Here $r_{500}$ is the radius of the halo where the average density within this radius is 500 times the critical density of the universe. } more spherical compared to non-radiative simulations. At small radii ($r/r_{500}< 0.1$), gas cooling leads to a more oblate gas shape due to the formation of  a rotating gaseous slab. In the non-dissipative simulations, on the other hand, the gas shape becomes more spherical in the central regions due to the enhanced random gas motions in the region with a flat entropy profile. Thus, by measuring the gas shape in the core, one can infer the level of gas motions in cluster cores.  By analyzing mock {\it Chandra} images of the simulated clusters, we showed that the baryonic effects on the ellipticity profiles of galaxy clusters should be discernible using X-ray observations.  

In this companion paper we compare a sample of local clusters observed by {\it Chandra} and {\it ROSAT} to the same  suite of hydrodynamical cluster simulations used in Paper I to constrain the fraction of gas that cooled to form the central galaxy in real clusters.  A similar comparison has recently been carried out by \citet{fang_etal09}. Our study is different from \citet{fang_etal09} in two important respects. First, we analyze and compare significantly larger samples of both observed and simulated clusters. Second, we focus on the effects of baryon condensation on the ellipticity of equipotential surfaces at larger radii, while \citet{fang_etal09} focused on the effects of more recent cooling in the central regions of clusters. 

This paper is organized as follows. We describe the simulations in Section~\ref{sec:sim} and the X-ray data in Section~\ref{sec:obs}. In Section~\ref{sec:results}, we present the comparison between the ellipticity profiles of observed and simulated clusters. We conclude with a summary and discussion in Section~\ref{sec:discussion}.


\section{Simulations}
\label{sec:sim}


\subsection{Simulation data}
\label{subsec:simdata}

We use high-resolution cosmological simulations of
16 cluster-sized systems in the flat concordance {$\Lambda$}CDM model:
$\Omega_{m}=1-\Omega_{\Lambda}=0.3$, $\Omega_{b}=0.04286$,
$h=0.7$ and $\sigma_8=0.9$, where the Hubble constant is defined as
$100h{\ \rm km\ s^{-1}\ Mpc^{-1}}$, and $\sigma_8$ is the power
spectrum normalization on an $8h^{-1}$~Mpc scale.  The simulations
were performed with the Adaptive Refinement Tree (ART)
$N$-body$+$gasdynamics code \citep{kra99,kra02}, an Eulerian code that
uses adaptive refinement in space and time, and (non-adaptive)
refinement in mass \citep{klypin_etal01} to reach the high dynamic
range required to resolve cores of halos formed in self-consistent
cosmological simulations. In order to assess the effects of gas cooling 
and star formation on the cluster shapes, we conducted each cluster simulation with two
different prescriptions for gas dynamics. In one set of runs we treated 
only the standard gas dynamics for the baryonic component without 
either radiative cooling or star formation.  We refer to these as 
non-radiative (NR) runs.  In the second set of runs, we included gas, 
metallicity-dependent radiative cooling, star formation, and supernova feedback (CSF).  
The simulations are discussed in detail in \citet{nagai_etal07a, nagai_etal07} 
and we refer the reader to these papers for more details. 
Here we summarize the main parameters of the simulations.

Simulations were run using a uniform 128$^3$ grid and eight levels
of mesh refinement in computational boxes of $120\,h^{-1}$~Mpc and 
$80\,h^{-1}$~Mpc on a side, respectively. These simulations achieve a dynamic range of $32,768$ and 
a formal peak resolution of $\approx 3.66\,h^{-1}$~kpc and $2.44\,h^{-1}$~kpc, corresponding to an 
actual resolution of $\approx 7\,h^{-1}$~kpc and $5\,h^{-1}$~kpc for the $120\,h^{-1}$~Mpc and
$80\,h^{-1}$~Mpc boxes, respectively.  Only regions of $\sim
3-10\,h^{-1}$~Mpc around each cluster was adaptively refined, while the rest of the
volume was followed on the uniform $128^3$ grid.  The mass resolution, 
$m_{\rm part}$, corresponds to an effective $512^3$ particles in the entire box, or
a Nyquist wavelength of $\lambda_{\rm Ny}=0.469\,h^{-1}$ and $0.312\,h^{-1}$
comoving Mpc for box sizes of $120\,h^{-1}$~Mpc and $80\,h^{-1}$~Mpc, respectively.  
These correspond to $0.018\,h^{-1}$ and $0.006\,h^{-1}$~Mpc in physical units at the 
initial redshifts of the simulations. The dark matter (DM) particle mass in the regions around each 
cluster was $9.1\times 10^{8}\,h^{-1}\, {M_{\odot}}$  and
$2.7\times 10^{8}\,h^{-1}\,{M_{\odot}}$ for the two box sizes, while other regions were
simulated with lower mass resolution. 


\subsection{Mock {\it Chandra} images}

To compare our simulation results to observations of X-ray clusters, 
we analyze mock {\it Chandra} images for each simulated cluster, with the instrumental 
response of real X-ray instruments taken into account, and measured the ellipticity profiles
of the gas emission in these mock images.  Below we give an overview of the methods 
used to generate the mock images.  Detailed descriptions of these mock images can be found 
in Section 3.1 of \cite{nagai_etal07a}. 

First we created X-ray flux maps of the simulated clusters viewed along 
three orthogonal projections. A flux map is computed by projecting 
the X-ray emission of hydrodynamic cells enclosed within three virial radii of
a cluster along the line of sight.  The X-ray emissivity in each
computational grid cell is computed as a function of proton and
electron densities, gas temperature and metal abundance. Emission from 
gas with temperature less than $10^5$~K is excluded as it is below the 
{\it Chandra} bandpass.  We then convolved the emission spectrum with
the response of the {\it Chandra} front-illuminated CCDs and draw a
number of photons at each position and spectral channel from the
corresponding Poisson distribution.  Each map has an exposure time of
100 ks (typical for deep observations) and includes a background with
the intensity corresponding to the quiescent background level in the
ACIS-I observations \citep{markevitch_etal03}.  The resolution of all
the maps is 6 kpc pixel$^{-1}$. 

From these data, we generated X-ray images in the 0.7--2 keV 
band and used them to identify and mask out all detectable 
small-scale clumps, as is routinely done in observational analyses. 
Our clump detection is based on a wavelet decomposition algorithm 
\citep{vikhlinin_etal98}. The holes left by masking out substructures in 
the photon map are filled in by the values from the decomposed map 
of the largest scale in wavelet analysis. We have tested that this method preserves 
the global shape of the photon distribution well. 
The background is removed when estimating ellipticities as it 
can bias them low at radii where background dominates the intrinsic emission. 
Throughout this paper we assume that the cluster redshift is 
$z_{\rm obs}=0.06$ for the $z=0$ sample.  

To investigate how the dynamical states of clusters affects their shapes, we divided our sample
into relaxed and unrelaxed sub-samples based on visual examinations of their mock X-ray images. 
Relaxed clusters are characterized by regular shapes and the absence of prominent substructures in all 
three of their orthogonal mock maps. Details of the classification can be found in \cite{nagai_etal07a}. 
We focused our comparison with observations on relaxed clusters in this paper. There are a total 
of 21 mock images in which the clusters are relaxed. 

We estimated axis ratios following \cite{dubinski_carlberg91} and \cite{kazantzidis_etal04} where we 
calculated the inertia tensors of the intracluster gas distribution within radial shells. This method provides 
a robust measure of shape \citep[see detailed discussion and tests by][]{zemp_etal11}. The ellipticities are 
measured by computing the inertia tensor of the photon distribution weighted by the photon counts and 
are estimated within radial {\em annular} bins. Details about the shape estimation method 
can be found in Sections~3.1 and 4.1 of Paper I and Section~\ref{subsec:methods} of the current paper.


\section{Observations}
\label{sec:obs}

\subsection{Data}
\label{subsec:data}

\begin{table*}[htbp]
\begin{center}
\caption{Properties of Observed Sample of $z<0.1$ Clusters}\label{tab:obs}
\begin{tabular}{l c c c c c  }
\hline
\hline
Cluster \hspace*{5mm} & {\em Chandra}(C)/{\em ROSAT}(R) &
{$z$} &
{$r_{500}$ (kpc)} & $-d\log\rho_{\rm gas}/d\log r$ at $0.04r_{500}$ & relaxed? \\
\hline
2A 0335 & C,R & 0.035 & 902.4 & 1.262 & yes\\
A 119 & C,R & 0.044 & 1113.9 & 0.287 & no\\
A 133 & C & 0.057 & 923.1 & 1.457 & yes\\
A 1644 & C & 0.047 & 1090.9 & 0.773 & no\\
A 1650 & C & 0.082 & 1109.1 & 0.715 & yes\\
A 1651 & C,R & 0.085 & 1193.5 & 0.605& yes\\
A 1736 & C & 0.045 &  916.9 & 0.019& no\\
A 2052 & C,R &0.034 & 831.9 & 1.025 & yes\\
A 2063 & C,R & 0.034 & 886.0 & 0.719 & yes\\
A 2065 & C & 0.072 & 1142.9  & 0.758 & no\\
A 2142 & C,R & 0.090 & 1519.3 & 0.664 & yes\\
A 2147 & C,R & 0.036 & 990.0 & 0.281 & no\\
A 2151 & C,R & 0.035 & 674.4 & 0.859 & no\\
A 2163 & C,R & 0.034 & 886.0 & 0.719 & yes \\
A 2199 & C,R & 0.030 & 954.0 & 0.802 & yes\\
A 2204 & C,R & 0.151 & 1373.9 & 1.365 & yes\\
A 2244 & C,R & 0.099 & 1143.1 & 0.599 & yes\\
A 2256 & C,R & 0.058 & 1328.7 & 0.183 & no\\
A 2589 & C,R & 0.041 & 847.3 & 0.549 & yes\\
A 2597 & C,R & 0.083 & 948.1 & 1.115 & yes\\
A 2634 & C,R & 0.031 & 820.1 & 0.487 & yes\\
A 2657 & C,R & 0.040 & 886.1 & 0.761 & yes\\
A 3112 & C,R & 0.076 & 1078.8 & 1.008 & yes\\
A 3158 & C,R & 0.058 & 1079.9 & 0.466 & yes\\
A 3266 & C,R & 0.060 & 1396.3 & 0.104 & no\\
A 3376 & C,R & 0.045 & 976.8 & 0.362 & no\\
A 3391 & C,R & 0.055 & 1072.8 & 0.456 & yes\\
A 3395 & C,R & 0.051 & 1076.4 & 0.002 & no\\
A 3558 & C,R & 0.047 & 1137.0 & 0.655 & no\\
A 3562 & C,R & 0.049 & 1003.3 & 0.488 & yes\\
A 3571 & C & 0.039 & 1219.1 & 0.601 & yes \\
A 3667 & C,R & 0.056 & 1303.3 & 0.188 & no\\
A 399 & C & 0.071 & 1228.1 & 0.312 & no\\
A 401 & C,R & 0.074 & 1364.1 & 0.223 & yes\\
A 4038 & C,R & 0.029 & 806.8 & 0.690 & yes\\
A 4059 & C,R & 0.049 & 952.6 & 0.588 & yes\\
A 496 & C,R & 0.033 & 976.6  & 1.064 & yes\\
A 576 & C & 0.040 & 901.6 & 0.697 & yes \\
A 85 & C,R & 0.056 & 1221.2 & 1.002 & yes\\
EXO 0422 & C & 0.038 & 781.1 & 1.102 & yes\\ 
Hydra-A & C,R & 0.055 & 953.5 & 1.239 & yes\\
MKW 3s & C,R & 0.045 & 865.3 & 0.909 & yes\\
RX J1504 & C  & 0.217 & 1364.6 & 1.400 & yes\\
S 1101 & C,R & 0.056 & 784.4 & 0.937 & yes\\
Zw Cl1215 & C,R & 0.077 & 1194.7 & 0.208 & yes\\
\hline
\end{tabular}
\end{center}
\end{table*}

\begin{figure}[t]
\begin{center}
\epsscale{1.0}\plotone{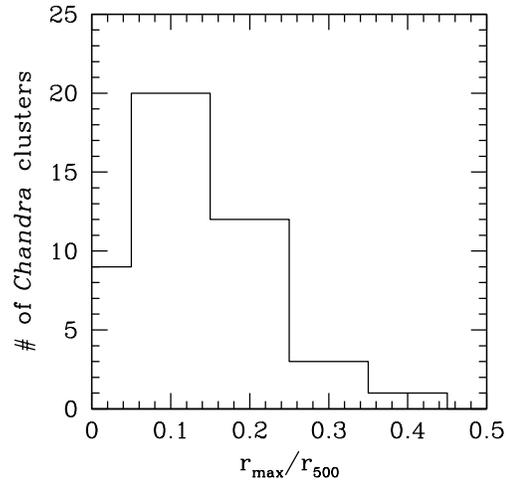}
\caption{Distribution of the shortest distance from the cluster center to 
the chip edge of our sample of {\it Chandra} clusters. 
}
\label{fig:rmax_hist}
\end{center}
\end{figure}

We analyze {\it Chandra} and {\it ROSAT}/PSPC X-ray observations of nearby galaxy clusters. 
The X-ray maps used are flat-fielded and with point source excluded. The cluster sample is taken 
from Table~2 of \cite{vikhlinin_etal09a}, and we refer the reader to this paper for a 
detailed discussion of selection and data reduction for this sample.  
Information on the cluster sample used in the current paper is also presented in Table~\ref{tab:obs}. 
The radius $r_{500}$ is derived from the best fit $M_{500}-Y_X$ relation from \cite{vikhlinin_etal09a}, 
where we assume a concordance ${\Lambda}$CDM model with  $\Omega_m=0.3,\Omega_{\Lambda}=0.7$, 
and Hubble parameter $h=0.72$. Clusters that exist in the {\it Chandra} or the {\it ROSAT} samples 
are labeled `C' or `R' respectively in the second column in Table~\ref{tab:obs}. 

We use both {\it Chandra} and {\it ROSAT} data whenever possible to probe different radial ranges within clusters.  
{\it Chandra} measurements provide exquisite cluster shape measurements in the inner regions of clusters.  While 
having a very good spatial resolution of $\sim4''$ (after binning to achieve good signal-to-noise), the ellipticity 
measurements are limited to within small radial range ($r\ll r_{500}$) due to the chip geometry of the ACIS camera 
on {\it Chandra}.  The centers of most of the {\it Chandra} clusters lie close to the edge of the ACIS-I chip, thus limiting 
the radial ranges available for robust shape measurement.  Figure~\ref{fig:rmax_hist} shows the distribution of the 
shortest distance from the cluster center to the closest edge of the ACIS-I chip, where we define the edge of the chip 
to be the pixels that have less than 40\% of the maximum value in the exposure maps. We further restrict our ellipticity 
measurements for each of the {\it Chandra} clusters to within 90\% of this distance. 

The large field of view (FOV) of {\it ROSAT} enables us to extend the ellipticity measurement to larger radii, 
although its spatial resolution is considerably worse than that of {\it Chandra}. The point-spread-function of 
{\it ROSAT}/PSPC has an FWHM  of $\sim 25 ''$. Hence we limit our {\it ROSAT} ellipticity measurement 
with the innermost radial bin being  greater than this value and only uses pixels which have fluxes above 
the background flux determined as the mean flux outside $r_{500}$ of the cluster. 

We also check our cluster sample for any bubbles or jet-like structures due to active galactic nuclei (AGNs) in the core region 
that could bias our ellipticity measurements. In general, these structures have little effect on the ellipticity, 
nevertheless we exclude the regions that contain these structures from our analyses wherever they are 
visually present in the images.

\subsection{Methods}
\label{subsec:methods}

We estimate the ellipticity of the cluster in an X-ray map by
computing the inertia tensor of the X-ray photon distribution
in annular bins. 
The inertia tensor is computed as
\begin{equation}
I_{ij} = \sum_{\alpha} {w_{\alpha}x_{\alpha,i} x_{\alpha,j}}, 
\label{eq:inertia_tensor}
\end{equation}
using pixels with photon count $w_{\alpha}$ if the ellipsoidal distance of the pixel
\begin{equation}
r_{\alpha} = \sqrt{x^{\prime\ 2}_{\alpha}+\left(\frac{y^{\prime}_{\alpha}}{1-\epsilon}\right)^2} 
\end{equation}
is within a spherical annulus $[R,R+\Delta R]$, where $R$ is the radius of the annulus .  
The primed coordinates are the coordinates of the pixel rotated to the 
frame of the principal axes of the inertia tensor. The ellipticity is initially set to zero, and
then computed iteratively as one minus the ratio of the square root of eigenvalues from 
diagonalizing the inertia tensor. The ellipticity, $\epsilon$, is defined as
\begin{equation}
\epsilon \equiv 1-\frac{b}{a},
\label{eq:ellipticity}
\end{equation}
where $a$ and $b$ are the semi-major and the semi-minor axis lengths from the fitted ellipse, 
respectively. The iteration stops when the ellipticity converges to within $1\%$ of its previous value.  

 Following Paper I, we identified point sources and substructures using wavelet algorithm \citep{vikhlinin_etal98}. 

As discussed in Section 2, the substructures are identified using wavelet algorithm (Vikhlinin et al 1998). In relaxed clusters only a few substructures are typically detected within $r_{500}$. The rest are mostly point sources identified as background AGNs. As it is difficult to distinguish AGNs from extended gas clumps within clusters at large radii (especially for {\it ROSAT} images), it is difficult to quantify the effects of clumps on the ellipticity measurements. At small radii, however, detectable clumps are rare and do not change our results except in a few obvious cases.  In order to robustly measure the global ICM shape while minimizing biases due to point sources and clumps, we mask them out and replace the holes with values from large-scale wavelet maps corresponding to scales greater than 32 pixels in each map. Improper treatment of point sources and clumps introduces significant differences in the ellipticity measurements. 

The radial binning is set up logarithmically within $r_{500}$ such that the photon count in each bin is approximately the same. In addition, we ensure that there are at least 25 pixels in each radial bin. Except for the innermost bin in a few cluster, all bins have $>100$ pixels. Finally, in our analysis, we allow the centroid to move for each annulus. Fixing the centroid will change the ellipticity typically by less than $10\%$.

We estimate errors on ellipticity measurements using a bootstrap approach similar to \cite{buote_etal02}, where we generate 100 realizations of the X-ray map and estimate the ellipticities using the method described above. The error is taken to be the dispersion of the bootstrap samples. The error is typically a few percent of the ellipticity value. 

We apply the same procedure above for our mock X-ray maps for consistent comparison with the observed clusters. 

\subsection{Dynamical state}
\label{subsec:dyn_state}

To characterize the dynamical state of each cluster, we classify our observational sample
into relaxed and unrelaxed clusters. This can be accomplished in two ways. The first one
is qualitative, where we classify clusters based upon visual examinations. If
their X-ray morphologies are regular and have no secondary peaks or filaments, we
classify them as relaxed, otherwise they are unrelaxed \cite[cf. Section 4.1.3 in][]{vikhlinin_etal09a}. 
This is the classification that we used to classify our mock images of simulated clusters in Paper I. 

The second method is more quantitative: we 
classify a cluster as relaxed, if the logarithmic slope of its gas density 
at $r=0.04r_{500}$ is less than $-0.5$ for the {\it Chandra} sample \citep{vikhlinin_etal07}. 
We were not able to apply this method to our CSF sample
because their gas density slopes are too steep at $r<0.1r_{500}$ due to ``overcooling.'' 
For a fair comparison between observation and simulations, we classify the cluster dynamical 
state visually. We have checked that the two methods result in similar samples of relaxed clusters, 
and our results for average ellipticity are within 1-$\sigma$ for the two observational samples defined 
in these ways. For the comparison with simulated cluster ellipticity profiles, we focus our comparison 
only on relaxed clusters because in this case ellipticity primarily reflects ICM physics rather 
than spurious effects due to transient substructure or incomplete relaxation of the gas distribution 
during and after mergers.


\section{Results}
\label{sec:results}


\begin{figure*}[t]
\begin{center}
\epsscale{1.0}\plotone{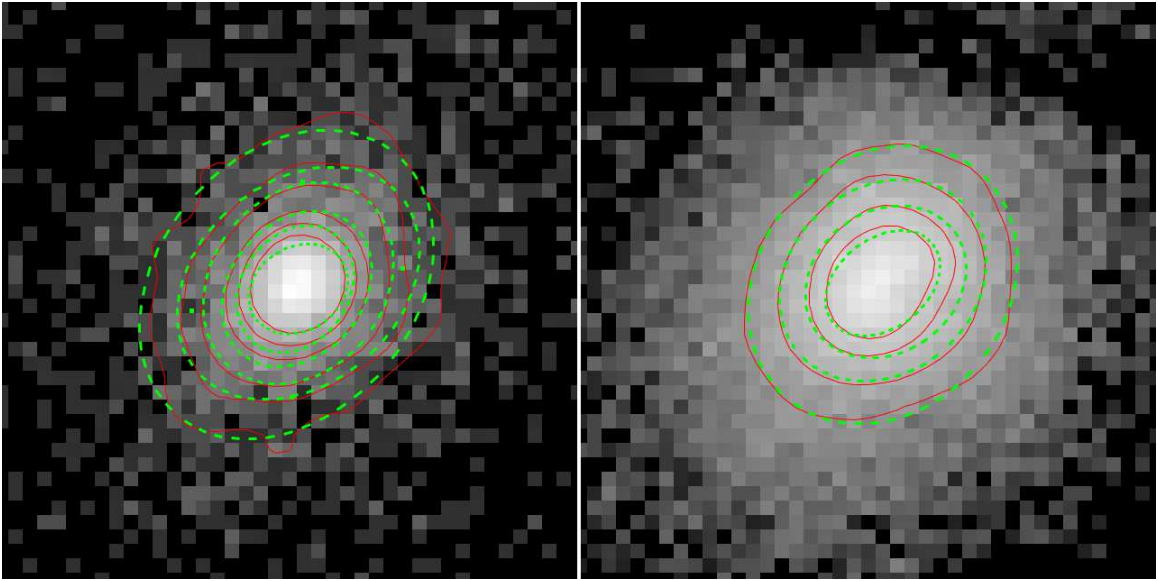}
\caption{X-ray image of A2597 observed by {\it ROSAT} ({left} panel) and {\it Chandra} ({right} panel. Shown in both panels are the X-ray isophotal contours (red thin lines) and the fitted ellipses (green dashed lines).  The pixel size is 5.97~kpc for the {\it Chandra} image and 22.7~kpc for the {\it ROSAT} image. The size of the major axis of the largest ellipse is $0.065 r_{500}$ in the {\it Chandra} image and $0.28 r_{500}$ in the {\it ROSAT} image, where $r_{500}=948$~kpc for this cluster. Note that the brightness scales of the two images are different. }
\label{fig:a2597}
\end{center}
\end{figure*}

\begin{figure*}[htbp]
\begin{center}
\epsscale{1.0}\plotone{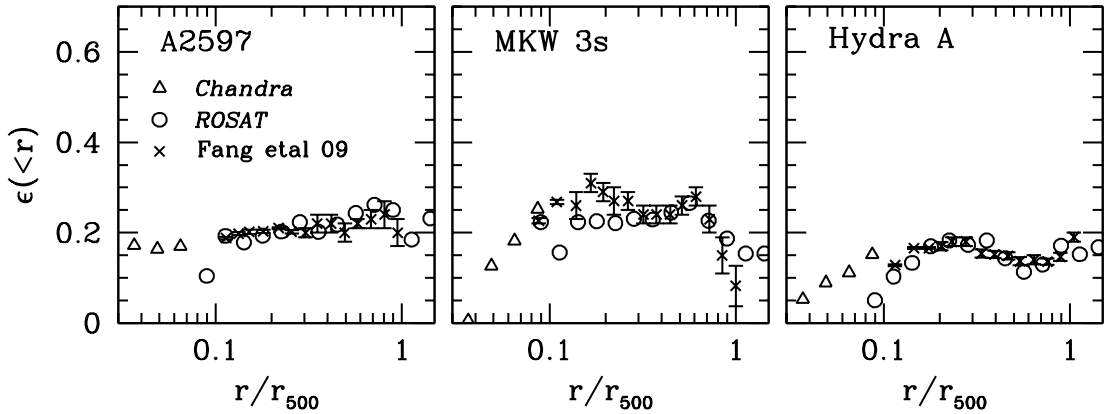}
\caption{Comparison of ellipticity profiles of three relaxed, cool-core clusters: A2597, MKW 3s and Hyrda A.  Here we compute the cumulative ellipticity using all pixels enclosed within a spherical region with radius $r$. Triangles and circles are the ellipticity measurement of our {\it Chandra} and {\it ROSAT} images, respectively. Crosses indicate the ellipticity measurements from \cite{fang_etal09}, which also combine Chandra data at small radii and {\it ROSAT}/PSPC observations at larger radii. Error bars are estimated using 100 bootstrap samples. 
The size of the circles 
indicates the size of the largest error bar.  
}
\label{fig:compare_F09}
\end{center}
\end{figure*}

\subsection{Test cases: A2597, MKW 3s, and Hydra A}

To demonstrate the robustness of our method applied to observed clusters, we first examine
the ellipticity measurements of three relaxed, cool-core clusters, A2597, MKW 3s, and 
Hydra A, for which independent cluster shape measurements exist in the recent literature 
\citep{fang_etal09}.  Note, however, that all three clusters exhibit AGN 
activity in their central regions.

As an illustration, we show in Figure~\ref{fig:a2597} the X-ray image of A2597 observed 
with {\it Chandra} and {\it ROSAT}. On each image, we show the isophotal contours and the fitted ellipse 
from our ellipticity estimation. There is a good agreement between the fitted ellipses and the X-ray contours, 
demonstrating the robustness of our method of estimating ellipticity on observed X-ray images of clusters.  
Note that there are X-ray cavities associated with AGN activity in the central region of A2597 
\citep{mcnamara_etal01}. However, structures associated with AGN activity are not discernible 
in most of our images because of lack of resolution, and therefore our fitted ellipses are not affected 
by these structures. 

Figure~\ref{fig:compare_F09} presents a comparison of the ellipticity profiles of A2597, MKW 3s, 
and Hydra A. The {\it triangles} and {\it circles} are the ellipticity measurement of our {\it Chandra} and 
{\it ROSAT} images respectively. {\it Crosses} are independent analyses of {\it Chandra} and {\it ROSAT}/PSPC 
observations by \cite{fang_etal09}.\footnote{Note that we have corrected for the radius used in \cite{fang_etal09} 
by normalizing the radial scale using the same $r_{500}$ reported in Table~\ref{tab:obs}.}  
Here, we computed the ellipticities using all pixels within a given radius 
rather than pixels in {\it annular} shells, in order to compare to the results 
of \cite{fang_etal09}.  
Figure~\ref{fig:compare_F09} shows that our ellipticity profile measurements are in good 
agreement with the shape measurements of \cite{fang_etal09}. 

\subsection{Comparison of Mean Ellipticity Profiles in Local X-Ray Clusters vs. $\Lambda$CDM Clusters}
\begin{figure*}[t]
\begin{center}
\epsscale{1.0}\plotone{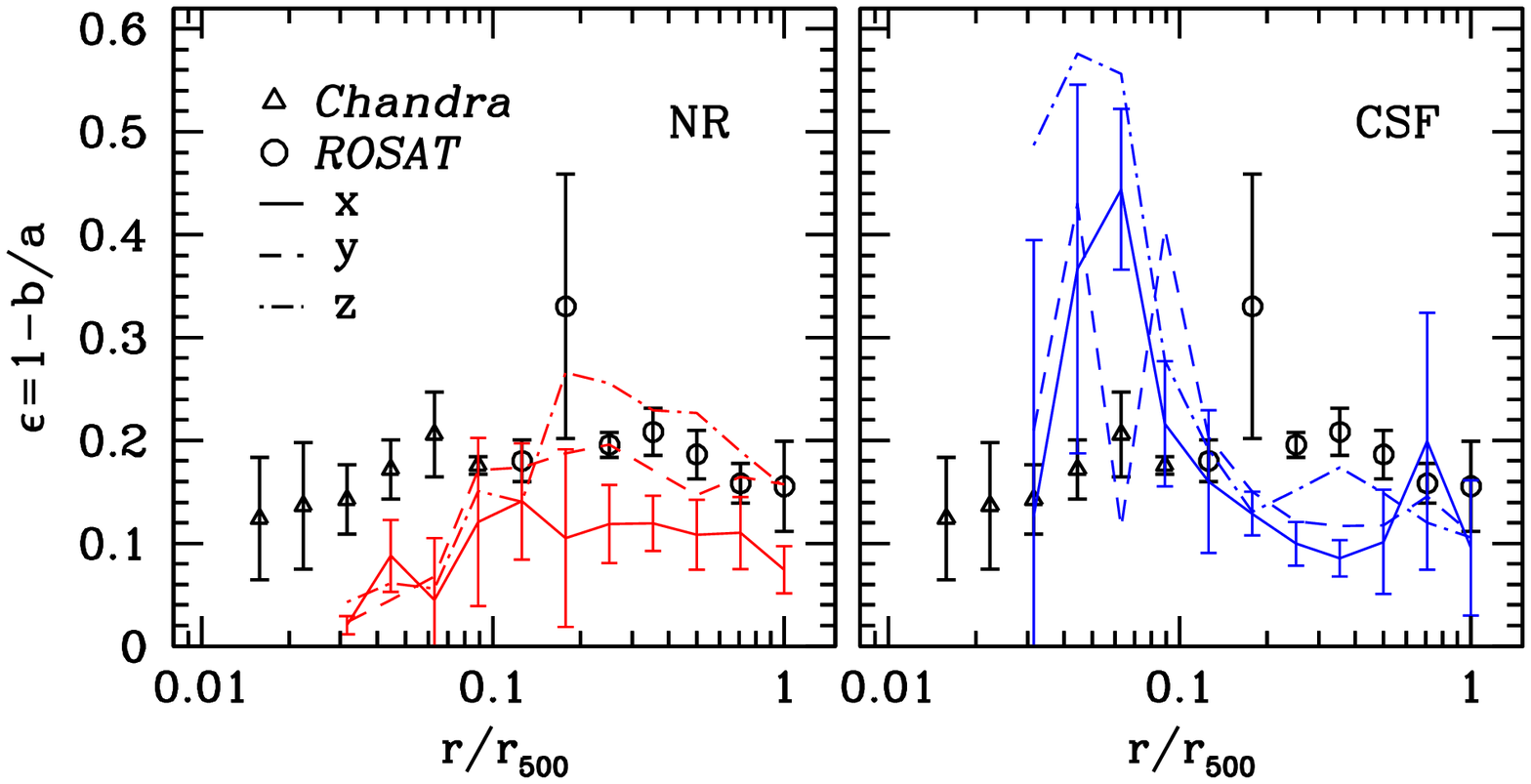}
\caption{Error-weighted average ellipticity profiles of the observed {\em relaxed} clusters shown as 
data points. Triangles show ellipticities estimated using the {\it Chandra} observations, while the 
circles are measurements from {\it ROSAT} data. 
The error bars show the jackknife errors on the error-weighted mean ellipticities. 
The left and right panels show comparisons between 
observed profiles and simulations without (NR) and with (CSF) cooling, respectively.  
The average ellipticity profiles of simulated clusters are plotted for the $x$, $y$ and 
$z$ projections, shown in solid, dashed, and dot-dashed lines, respectively. The error bars indicate the 
jackknife error of the mean ellipticity values for the $x$ projection. The errors for the other two 
projections are similar.} 
\label{fig:real_ellipticity}
\end{center}
\end{figure*}

\begin{figure*}[htbp]
\begin{center}
\epsscale{1.0}\plotone{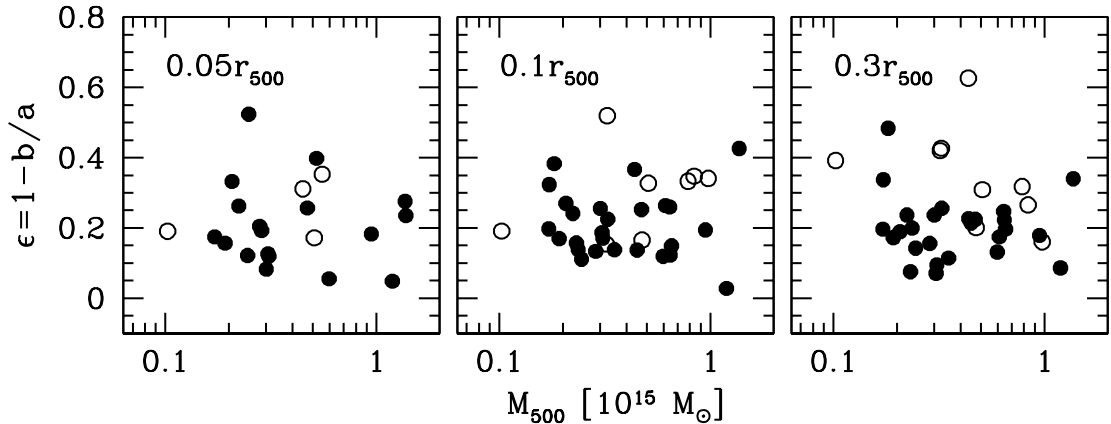}
\caption{
Ellipticity as a function of $M_{500}$ measured at $r=0.05r_{500}$ (left panel), $r=0.1r_{500}$ (middle panel), and $r=0.3r_{500}$ (right panel). Solid and open circles represent relaxed and unrelaxed clusters, respectively. Errors are not shown as they are typically much smaller than the ellipticity values. 
} 
\label{fig:ellipm}
\end{center}
\end{figure*}

\begin{table}
\begin{center}
\caption{Mean Ellipticity of the $z<0.1$ {\it Chandra} and {\it ROSAT} Clusters}
\label{tab:ellipticity}
\begin{tabular}{ c | c c c }
\hline
\hline
$r/r_{500}=$&$0.05$&$0.1$&$0.3$\\
\hline
Relaxed&$0.17\pm0.03$&$0.18\pm0.03$&$0.20\pm0.02$\\
\hline
Unrelaxed&$0.38\pm0.20$&$0.28\pm0.06$&$0.23\pm0.09$\\
\hline
All&$0.19\pm0.06$&$0.19\pm0.03$&$0.21\pm0.03$\\
\hline
\end{tabular}
\end{center}
\end{table}

Figure~\ref{fig:real_ellipticity} shows the average ellipticity profile for the 31 relaxed
clusters with both {\it Chandra} and {\it ROSAT} data (see Table~\ref{tab:obs}).   
The mean ellipticity of observed clusters (defined in Equation (~\ref{eq:ellipticity})) is $\epsilon\approx 0.2\pm 0.1$ 
across the radial range $0.05\leq r/r_{500} \leq 1$.  The error bars show the jackknife errors on the error-weighted 
mean ellipticities. 

In the same figure we compare the observed mean ellipticity profile to the profiles 
of the relaxed NR and CSF clusters, shown in the left and right panels, respectively. 
Note that for the simulated clusters, we do not report ellipticity for $r\lesssim 0.03r_{500}$ because it 
reaches the limit of the spatial resolution. 
The simulated ellipticity profiles are shown separately for each $x,y,z$ projection. Note that the $x,y,z$ axes 
correspond to the coordinates of the simulation box, and the projection is done along each orthogonal axis. Since the 
orientation of the cluster is arbitrary with respect to the box, each projection is a random sight line through each cluster. 

The number of relaxed clusters in our simulated sample is fairly small, so we also show the ellipticity 
profiles averaged over $x$, $y$ and $z$ projections independently (shown as solid, dashed, 
and dot-dashed lines, respectively) for both NR and CSF runs. This illustrates that the general
trends seen in the NR and CSF are not significantly affected by orientation effects due to small number statistics, 
as similar trends are seen in each independent projection.

In the core regions ($r\lesssim 0.1r_{500}$), the NR ellipticities describe the observations relatively well, with NR 
ellipticities close to $\epsilon_{\rm NR}  \approx 0.2$. On the other hand, the CSF ellipticities are significantly higher 
than the observed ellipticities in the core regions, with ellipticities of $\epsilon_{\rm CSF} \approx 0.3-0.5$ due to the 
cold gas slab forming in the center via a cooling flow. 

At larger radii, $0.1\lesssim r/r_{500} \lesssim 1$, the ellipticity profiles of the NR clusters continue to match the observed 
ellipticity profiles quite well with $\epsilon \approx \epsilon_{\rm NR} \sim 0.2$, although there is a hint that 
the simulated NR clusters may have slightly lower values of $\epsilon$ than the observed clusters.  On the other hand, 
the ellipticity profiles in the CSF simulation clusters are considerably lower than the observed ellipticities over this range 
of radii.  For example, at $r=0.2r_{500}$, $\epsilon_{\rm CSF} \approx 0.13$, compared to  $\epsilon \approx 0.22$ for the 
observed clusters. As discussed in the introduction, the differences between the ellipticities of the NR and CSF clusters 
at large radii reflect the more spherical shapes of the overall cluster potentials, 
and correspondingly more spherical shapes of the ICM in the CSF simulations, 
due to the formation of the massive central galaxies in those clusters. Observations thus clearly indicate that 
the central galaxy in the CSF simulations is too massive compared to observations. This provides independent 
evidence for overcooling in the CSF simulations.  

To investigate the dependence of ellipticity on cluster mass, we show cluster ellipticities measured at  
$r/r_{500} = 0.05,0.1, and 0.3$ for the observed clusters as a function of $M_{500}$  in Figure~\ref{fig:ellipm}. 
The average ellipticity at $r=0.05r_{500}$ is derived from the {\it Chandra} observations, while 
the ellipticity values at the larger radii are derived from the {\it ROSAT} sample. 
There is no obvious dependence of ellipticity on cluster mass, 
consistent with the simulation results presented in our Paper I.  The scatter in ellipticity among clusters 
is also similar across all radii considered. A weak trend of ellipticity with cluster mass may be present, 
but this sample of clusters is too small to constrain it robustly.

\begin{figure*}[t]
\begin{center}
\epsscale{1.0}\plotone{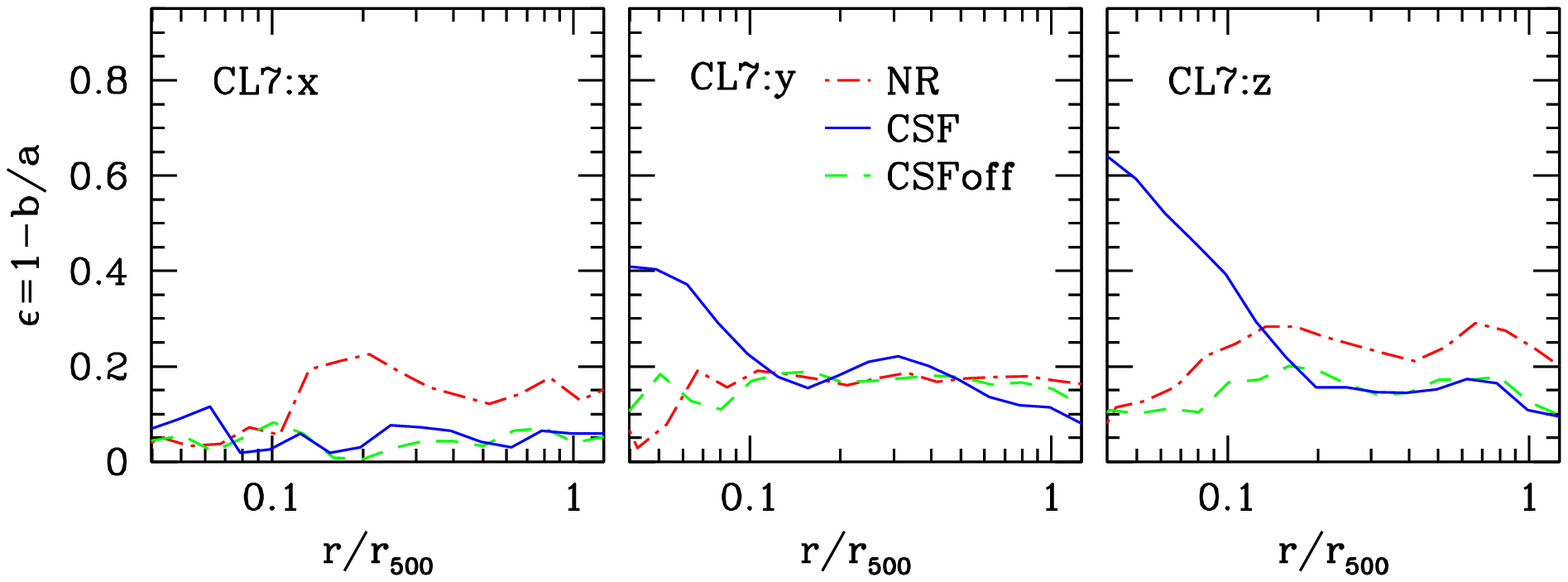}
\caption{ The ellipticity profiles for the three orthogonal projections of CL7 (left, middle, and right panels for the $x$,$y$, and $z$ projections, respectively) for the three runs with different physics. The red (dot-dashed), blue (solid), and green (dashed) lines represent the ellipticity profiles for the NR, CSF, and CSFoff (cooling and star formation turned off at redshift $z\leq2$) runs respectively. 
}
\label{fig:CL7}
\end{center}
\end{figure*}

\subsection{Effects of Dissipation on Ellipticity Profile}
\label{sec:CL7}

To investigate the dependence of the ellipticity on baryon dissipation, we re-simulated one of the 
clusters in simulation sample, CL7, with gas cooling, star formation, and
stellar feedback turned off at $z \leq 2.0$ (CSFoff).  CL7 is an isolated, relaxed cluster 
at $z=0$, which did not experience a major merger for several Gyr prior to $z=0$.  
The CSFoff experiment resulted in a final cluster with a stellar mass within 
$r=50$ kpc  about 2 times smaller than the stellar mass in the corresponding CSF simulation. 

We estimate the ellipticity for CL7 using the same methods that we 
applied to the observed clusters.  Figure~\ref{fig:CL7} shows the 
ellipticity profiles for the three orthogonal projections of CL7 
(left, middle, and right panels for the $x$,$y$, and $z$ projections, respectively) for 
the three runs. The red (dot-dashed), green (dashed) and blue 
(solid) lines represent the ellipticity profiles for the NR, CSFoff and CSF runs respectively. 
The agreement between the CSF and CSFoff profiles for $r\geq 0.1r_{500}$ shows that the effect of 
baryon condensation on the shape of the potential and ICM does not scale linearly with stellar mass 
within $r_{500}$.  Evaluation of this dependence will require further 
investigation with a much larger sample of clusters. 

Note that for $r<0.1r_{500}$ there is a
significant difference between CSF and CSFoff runs. 
The absence of cooling after $z=2$ prevented the formation of a cooling flow 
and extended flattened slab in the central gas distribution.  The absence of 
this gaseous slab resulted in ellipticity values at $r<0.1r_{500}$ closer to the values 
in the outer regions of the cluster. This clearly illustrates that the oblate slab 
in the core of this cluster in the CSF run is due to cooling during 
later stages of cluster evolution. 


\section{Summary and Discussion}
\label{sec:discussion}


In this study, we have compared mean ellipticity profiles of a sample of clusters simulated 
in the concordance $\Lambda$CDM cosmology with ellipticity profiles measured with 
{\it Chandra} and {\it ROSAT} X-ray observations of galaxy clusters.  

Our comparison presents several improvements over the previous work. In terms of analysis method, 
we have used mock X-ray photon maps of simulated clusters and used the same analysis pipeline on 
both the mock and real maps. This enabled more direct and consistent comparison between simulations 
and observations. In terms of sample size, we have significantly expanded the size of observational 
sample by analyzing 31 local clusters compared to 2 \citep{flores_etal07} and 9 \citep{fang_etal09}.  
Using this sample, we measured the average ellipticity of local relaxed clusters to be $\epsilon= 0.18 \pm 0.05$ 
in the radial range of $0.04\leq r/r_{500}\leq 1$.  Our results indicate that the shapes of X-ray emitting gas in 
galaxy clusters, especially at large radii, can be used to place constraints on cluster gas physics, and are 
potential probes of the history of baryonic cooling in galaxy clusters. Key results of the paper are 
summarized below. 

We showed that clusters formed in simulations with gas cooling and star formation (the CSF runs) generally 
predict ellipticities considerably smaller (i.e., a rounder ICM) than the observed clusters at radii 
$0.1\lesssim r/r_{500}\lesssim 1$, but they predict much higher ellipticities than the observed clusters in the 
core region ($r \lesssim 0.1r_{500}$).  The non-radiative simulations, on the other hand, have ellipticity profiles 
much closer to the observed cluster ellipticities in the range $r \gtrsim 0.05r_{500}$.   

The amount of baryonic dissipation occurring during cluster formation controls the stellar masses of central cluster 
galaxies.   The stellar masses of central cluster galaxies, in turn, control the shapes of the global potentials and 
the ICM of clusters at large cluster-centric radii. Our comparison therefore suggests that the masses of central galaxies 
in observed clusters are much smaller than the corresponding masses in our CSF simulations with gas cooling 
and star formation. This is a manifestation and an independent confirmation of the ``overcooling'' commonly seen 
in cluster formation simulations with radiative cooling and star formation \citep[e.g.,][]{borgani_kravtsov09}.  To probe 
the degree to which central galaxy mass is overestimated we have re-simulated one of our relaxed clusters with cooling 
and star formation turned off at $z \leq 2$. The cluster in this re-simulation had a considerably smaller final stellar mass, 
but still exhibited an ellipticity profile consistent with the CSF simulations at large radii.  This indicates that the sensitivity 
of the ICM ellipticity to the mass of central galaxy is nonlinear and will need to be quantified using larger samples of objects. 

In the central regions of clusters, $r\lesssim 0.1r_{500}$, the ICM shapes in clusters formed in simulations with cooling 
are generally oblate due to cooling flows that develop in the central regions and lead to the formation of rotating thick 
gas slabs.  This result is consistent with the results of \citet{fang_etal09}, who analyzed a subset of the cluster sample 
we study in this paper.  However, these authors found the sizes of the oblate slabs to be about a factor of two larger than 
we find here \citep[see discussion in][]{lau_etal11}.  The formation of such cooling flows and slabs occurs during late 
stages of cluster evolution.  For instance, in the CSFoff re-simulation, with cooling and star formation turned off at $z\leq2$, 
the shape of the gas distribution near the cluster center is not oblate and the cluster ellipticity is nearly constant with radius. 

The ICM shapes in the central and outer regions of clusters therefore probe and constrain \emph{different} regimes of 
baryonic cooling. The shapes at large radii are sensitive to the overall amount of gas that has cooled and condensed to 
form a central cluster galaxy throughout the entire cluster evolution.  We reiterate that our results indicate that the relationship 
between ICM shapes at large radii and the amount of gas that has cooled to form the central cluster galaxy is nonlinear and 
may be more well calibrated with a larger simulation set.  Alternatively, ICM ellipticities in the central regions of clusters probe 
recent and ongoing gas cooling. 

As discussed in Paper I, random, isotropic gas motions in the cluster cores result in more 
spherical gas shapes in the central regions of non-radiative clusters.  This leads to a downturn in the ellipticity 
profiles toward more spherical ellipticity values at $r/r_{500}\lesssim 0.1$ (see the left panel of 
Figure~\ref{fig:real_ellipticity}).  The mean  ellipticity profiles of observed clusters do not exhibit such 
a downturn at these radii. This implies that real clusters do not have gas motions as significant 
as those in our NR simulations. The likely reason for this difference is that observed, relaxed clusters have 
monotonically decreasing entropy profiles \citep[e.g.,][]{cavagnolo_etal09}, while the simulated NR clusters 
have flat entropy profiles at $0.15 \lesssim r/r_{500} \lesssim 0.2$ which are more conducive to convective 
random gas motions. 

A recent work by \citet{biffi_etal11} also studied the effects of baryon cooling on gas ellipticity and velocity structure 
using GADGET SPH code. While our results based on the AMR grid code are consistent with the results of \citet{biffi_etal11} 
in the outer regions of clusters, there are noticeable differences in the central regions.  Specifically, the cooling
runs performed by the SPH code do not produce a rotationally supported gas disk in the central regions of 
relaxed clusters, as seen in the AMR simulations. One possible explanation for such disparate results is the 
difference in disruption of gas clumps in SPH and AMR simulations \citep[e.g.,][]{agertz_etal07}. Gas clumps 
in the SPH code do not disrupt as efficiently as those in the AMR code due to SPH's poor ability to capture fluid instabilities 
(Kelvin--Helmholtz in particular). Since the unstripped gas loses angular momentum while the stripped gas does not, 
rotationally supported gas disk does not build up in the core of SPH clusters as extended as that in the AMR clusters.

The ellipticity profiles of clusters in non-radiative simulations agree with observed ellipticities well, which is consistent 
with results of \citet{fang_etal09}, who used a smaller sample of clusters with {\it Chandra} and {\it ROSAT}/PSPC data 
to derive mean ellipticity profile.  Similar results were also found by \cite{flores_etal07}, who used a subset of our 
non-radiative cluster simulation and showed that their shapes are in general agreement with observations.  
Our results are also consistent with the results of \citet{kawahara10}, who showed that distribution of cluster ellipticities 
derived from {\it XMM-Newton} observations are consistent with the predictions of dissipationless $N$-body simulations 
of clusters by \citet{jing_suto02}.  The broad accord between our results and these previous studies suggests that the 
ellipticity profiles in clusters formed in non-radiative simulations are in reasonable agreement with the ellipticity profiles 
of observed clusters. 

One possible concern with the comparison between simulations and observations presented in this paper is the possible 
bias in the ellipticities of simulated clusters due to the fact that our cluster sample was simulated assuming $\sigma_8 = 0.9$, 
while the most recent estimates indicate $\sigma_8 = 0.80\pm 0.02$ \citep[e.g.,][]{vikhlinin_etal09,rozo_etal10,jarosik_etal11}. 
The effect of a lower $\sigma_8$ is that halos form later, leading to higher ellipticities \citep{allgood_etal06}.  
This effect is likely to be small, so that it would not change our qualitative conclusions.  
For example, \citet{maccio_etal08} show that effect of changing $\sigma_8$ from $0.9$ to $0.8$ changes average 
minor-to-major axis ratios, $c/a$, of halos by only $\approx 0.03$.  This systematic shift is considerably smaller than the 
differences discussed in the present paper.  Furthermore, the sense of this effect would be to bring the NR simulations into 
better accord with the observed cluster sample.

On the observational side, a possible concern is that buoyant bubbles or jets associated with the central AGN activity could 
modify the morphology of gas in the core regions, resulting in higher observed ellipticities. We have visually examined our 
observed sample for any signatures of AGN activity and have excluded regions that seem to be affected by AGNs. However, 
it is still possible that some regions influenced by AGN activity are unresolved in our current sample, or that buoyant bubbles 
have had an impact on ICM morphology, but subsequently subsided so that they are not observed at present.

In this paper we compared the shapes of simulated $\Lambda$CDM clusters to the ICM observed in a local X-ray cluster sample. 
A natural extension of this work is to probe the ICM shape of higher redshift clusters with deeper X-ray imaging and high-resolution 
Sunyaev--Zel'dovich (SZ) observations. It would also be useful to carry out such comparisons with larger samples of simulated clusters 
for a range of cooling and feedback models, resulting in a wide and well-sampled range of central galaxy stellar masses. 
High-resolution SZ imaging is a potentially powerful means of extending this analysis to higher redshifts, because of the redshift 
independence of the SZ effect.  Several high-resolution SZ imaging experiments are in operation or under construction, including 
ALMA\footnote{Atacama Large Millimeter/submillimeter Array}, CARMA\footnote{Combined Array for Research in Millimeter-wave 
Astronomy}, CCAT\footnote{Cornell Caltech Atacama Telescope}, and MUSTANG\footnote{MUltiplexed Squid TES Array at Ninety GHz}. 
Detailed comparisons between SZ shapes in simulations and observations should provide further tests of the baryon dissipation in 
galaxy clusters over a wide range of redshifts. Such comparisons should inform theories of the formation of galaxies and galaxy 
clusters and may play a complementary role in efforts to refine constraints on the concordance cosmological model.

\acknowledgments 
The cosmological simulations used in this study were performed on the IBM RS/6000 SP4 system (copper) at the National Center 
for Supercomputing Applications (NCSA). This work was supported in part by the facilities and staff of the Yale University Faculty of 
Arts and Sciences High Performance Computing Center. EL and DN acknowledge the support from the NSF grant AST-1009811, by 
NASA ATP grant NNX11AE07G, and by Yale University.  
AK was supported by NSF grants AST-0507596 and AST-0807444, NASA grant
NAG5-13274, and by the Kavli Institute for Cosmological Physics at the University of Chicago 
through grants NSF PHY-0551142 and PHY-1125897. ARZ was supported by the Pittsburgh 
Particle physics, Astrophysics, and Cosmology Center (PITT PACC) and by the 
NSF through grants AST-0806367 and AST-1108802.  

\bibliographystyle{hapj}
\bibliography{ms}

\end{document}